\documentclass[12pt]{article}
\usepackage{cgpg}
\usepackage{graphicx}
\usepackage{psfrag}

\makeatletter

\renewcommand{\ps@plain}{%
        \renewcommand{\@evenhead}{}%
        \renewcommand{\@oddhead}{}%
        \renewcommand{\@evenfoot}{\hfil\small{\textbf\thepage}\hfil}%
        \renewcommand{\@oddfoot}{\@evenfoot}}%

\renewcommand\section{\@startsection {section}{1}{\z@}%
                                   {-3.5ex \@plus -1ex \@minus -.2ex}%
                                   {2.3ex \@plus.2ex}%
                                   {\reset@font\bfseries}}
\renewcommand\subsection{\@startsection{subsection}{2}{\z@}%
                                     {-3.25ex\@plus -1ex \@minus -.2ex}%
                                     {1.5ex \@plus .2ex}%
                                     {\reset@font\itshape}}
\newlength{\capsize}
\renewcommand{\@makecaption}[2]{%
  \vskip\abovecaptionskip
  \sbox\@tempboxa{\small{\bfseries #1}\/: #2}%
  \ifdim \wd\@tempboxa >\capsize
   {\advance\leftskip by 0.1\textwidth \advance\rightskip by 0.1\textwidth
    {\small {\bfseries #1}: #2}\par}
  \else
    \hbox to\hsize{\hfil\box\@tempboxa\hfil}%
  \fi
  \vskip\belowcaptionskip}

\makeatother

\newtheorem{defn}{Definition}
\newcommand{\goesto}{\rightarrow}

\title{\large \textbf{Causality in Spin Foam Models}}

\author{%
    \normalsize \textbf{Sameer Gupta}%
    \thanks{Email address: gupta@gravity.phys.psu.edu}\\
    \normalsize{\it Center for Gravitational Physics and Geometry, Physics
                Department}\\
    \normalsize{\it The Pennsylvania State University}\\
    \normalsize{\it 104 Davey Laboratory, University Park, PA 16802, USA}}

\date{\normalsize{August 5, 1999}}

\AccentVectors

\begin{document}

\maketitle

\begin{abstract}
We compute Teitelboim's causal propagator in the context of canonical loop
quantum gravity. For the Lorentzian signature, we find that the resultant power
series can be expressed as a sum over branched, colored two-surfaces with an
intrinsic causal structure. This leads us to define a general structure which
we call a ``causal spin foam''. We also demonstrate that the causal evolution
models for spin networks fall in the general class of causal spin foams.
\end{abstract}

\section{Introduction}

In any theory of quantum gravity, classical space-time is likely to arise only
as an approximate concept. This problem is especially obvious in canonical
quantum gravity. In particular, there is no notion of time in quantum gravity.
Hence, it may seem like the causal structure of the classical metric is absent
in the quantum theory as well.

However, we can incorporate the causal structure at the fundamental level in a
theory of quantum gravity. In a path-integral formulation of quantum gravity,
we express the transition amplitude from one quantum three-geometry to another
as a sum of histories. Causality can be included by limiting the sum to only
those histories which have a well-defined causal structure. This has been
suggested as early as \cite{Teitel83} by Teitelboim. There has been further
work in this direction in mini-superspace models \cite{Louko91}. Similar ideas
have also been explored in the work on causal evolution of spin networks
\cite{MarkoSmo97,Marko97} and in two space-time dimensions \cite{AmbLoll98}. In
this paper, we explore some of these ideas more concretely in the context of
loop quantum gravity. We shall construct a formal power series expansion of
Teitelboim's causal propagator in canonical (loop) quantum gravity. We show
that each term in the power series has a fixed causal structure which suggests
that each of the terms may admit an interpretation as a quantum Lorentzian
space-time. We then define a more general object --- a ``causal spin foam'' ---
which incorporates the key features which allow us to do so.

Causal spin foams are an extension of spin foam models which have been
introduced in the quantization of Euclidean gravity \cite{ReisRov97, SpinFoams,
Baez98}. These are space-time models which have a close connection with the
loop approach to canonical quantum gravity. We shall also relate causal spin
foams to the work on causal evolution of spin networks which was proposed as a
model for Lorentzian quantum gravity.

The rest of this paper is organized in the following manner. We begin with a
brief introduction to spin networks and canonical loop quantum gravity,
spelling out the main ideas that we use later. Then, in section 3, we give the
details of the construction of the propagator. Following that, we give the
general definition of a causal spin foam and show how the causal evolution
model relates to it. Finally, we end the paper with a discussion of some open
and interesting issues.

\section{Spin Networks in Canonical Quantum Gravity}

Canonical general relativity (GR) can be written as a theory of a real SU(2)
connection $A_a^i$ over a compact three-manifold $\Sigma$ \cite{Ashtekar86,
Barbero94}. The conjugate variable to the connection is the densitized triad
$\tilde{E}_i^b$ which takes values in the lie algebra \underline{su}(2). Using
these variables, the constraints of GR are three Gauss constraints which impose
SU(2) invariance, three diffeomorphism constraints which impose diffeomorphism
invariance on $\Sigma$, and the Hamiltonian constraint, which is the generator
of coordinate time evolution. A review with the details of and references for
the construction of the quantum theory can be found in \cite{Rovelli98}. In
this section, we present a brief summary of the key results we shall use in
this paper.

\subsection{The Kinematical Hilbert Space}

The kinematical Hilbert space of SU(2) invariant states, ${\cal
H}_\mathrm{kin}$, has a convenient orthonormal basis called the spin network
basis \cite{Spinnet}. A spin network state is constructed as follows: consider
an oriented, closed graph $\gamma$ which is embedded in the spatial manifold
$\Sigma$. To each edge of $\gamma$ assign a representation $j_e$ of SU(2) and
to each node of the graph, assign an intertwiner $I_n$ such that
\[ I_n: \bigotimes_\mathrm{incoming\ edges} j_e \longrightarrow
\bigotimes_\mathrm{outgoing\ edges} j_e\ .\] A spin network is then defined by
the triplet $S = (\gamma, \vec{\jmath}, \vec{I})$. A spin network state is then
defined as
\begin{equation}
    \psi_S(A) = \prod_e \prod_n I_n R^{(j_e)}(U(e,A))
\end{equation}
where $R^{(j_e)}(U(e,A))$ is the holonomy of the connection $A$ along the edge
$e$ in the representation $j_e$ of SU(2). The inner product between two spin
network states $\psi_S$ and $\psi_S'$ is given by
\begin{equation}
    \iprod{\psi_S}{\psi_{S'}} =
    \delta_{\gamma,\gamma'}\delta_{\vec{\jmath},\vec{\jmath}'}%
    \delta_{\vec{I},\vec{I}'}
\end{equation}

In this basis, geometric operators such as area and volume are diagonal and
have discrete spectra (This has been discussed by a number of authors. See for
example references \cite{Q-geom}). Spin network states can thus be thought of
as describing discrete three-geometries.

\subsection{The Diffeomorphism Constraint}

    The group of spatial diffeomorphisms acts on the states in ${\cal
H}_\mathrm{kin}$ in a natural fashion. The action of a finite diffeomorphism
$\phi$ on a spin network state for example is given by
\begin{equation}
    \hat{U}(\phi) \psi_S(A) = \psi_S(\phi^{-1}A) = \psi_{\phi\circ S}(A).
\end{equation}
This simply means that the action of the diffeomorphism is to send the state
to one which is based on the shifted graph.

    The space of solutions of the diffeomorphism constraint of the theory,
${\cal H}_\mathrm{diff}$, is defined by $s$-knot states which are
diffeomorphism equivalence classes of spin networks. ${\cal H}_\mathrm{diff}$
is a subset of the algebraic dual of ${\cal H}_\mathrm{kin}$, so every
$s$-knot state acts on spin networks
\begin{equation}
    \braket{s}{S} \not= 0 \quad \mathrm{iff} \quad s\in\{S\}
\end{equation}
where $\{S\}$ is the diffeomorphism equivalence class to which $\ket{s}$
belongs. We can use this to define an inner product on diffeomorphism invariant
states:
\begin{equation}\label{diffip}
    \braket{s}{s'} = \braket{s}{S'}
\end{equation}
where $\ket{S'}$ is a representative of the diffeomorphism equivalence class
defined by $\ket{s'}$.

\subsection{The Hamiltonian Constraint}

The Hamiltonian constraint is the least understood piece of the dynamics of
canonical GR. One of the reasons for this is that it has no obvious geometrical
interpretation like the Gauss and diffeomorphism constraints. The smeared
Hamiltonian
\begin{equation}\label{ham-evol}
    \hat{H}[N] = \int d^3x\, N(x,t)hat{H}(x),
\end{equation}
where $N(x,t)$ is the lapse function and $\hat{H}(x)$ is the Hamiltonian
constraint, can be thought as the generator of coordinate evolution in
canonical gravity.

\begin{figure}[t]
    \begin{center}
        \psfrag{Amp}{$A^{+-}_x(s)$}
        \psfrag{Amp1}{$B^{+-+-}_x(s)$}
        \psfrag{x}{\small $x$}
        \psfrag{y}{\small $y$}
        \psfrag{yp}{\small $y'$}
        \psfrag{ypp}{\small $y''$}
        \psfrag{i}{\small $i$}
        \psfrag{j}{\small $j$}
        \psfrag{k}{\small $k$}
        \psfrag{jminus1}{\small $j-1$}
        \psfrag{iplus1}{\small $i+1$}
        \psfrag{1}{\small $1$}
        \psfrag{kminus1}{\small $k-1$}
        \psfrag{a}{(a)}
        \psfrag{b}{(b)}
        \psfrag{C}{\small $\hat{C}_E$}
        \psfrag{T}{\small $\hat{T}$}
        \makebox[\columnwidth]{
            \includegraphics[height=1.25in,keepaspectratio,clip]{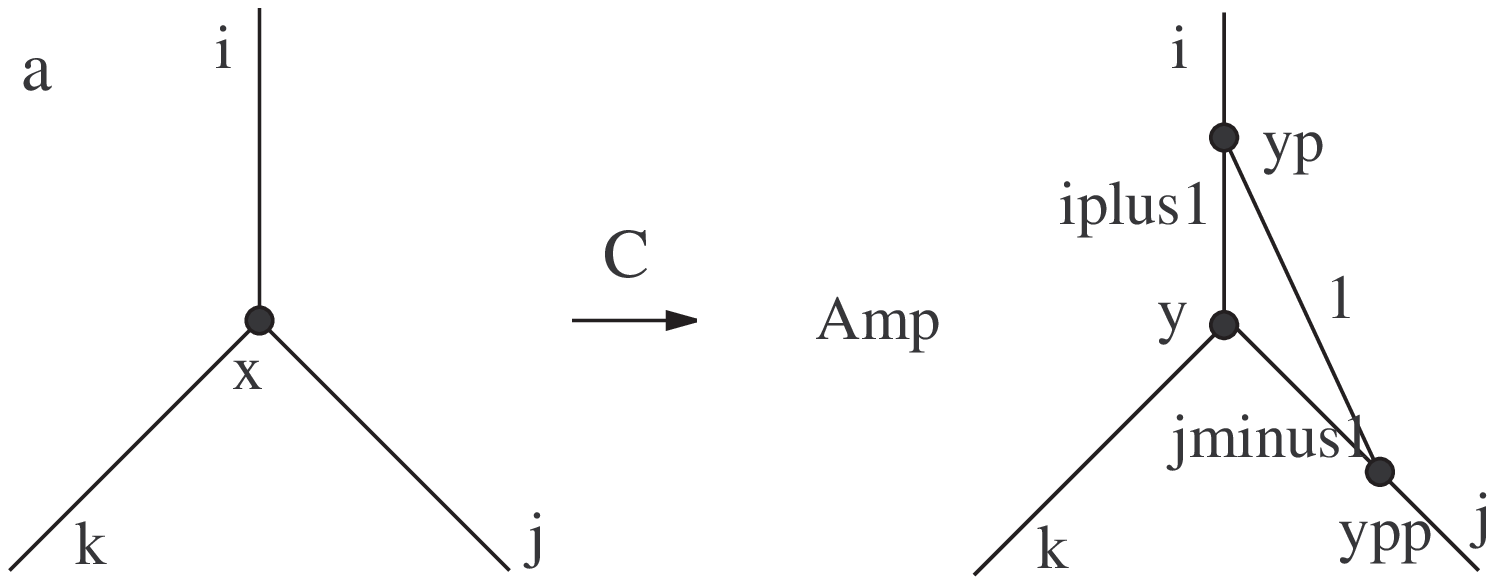}
            \hspace*{0.25in}
            \includegraphics[height=1.25in,keepaspectratio,clip]{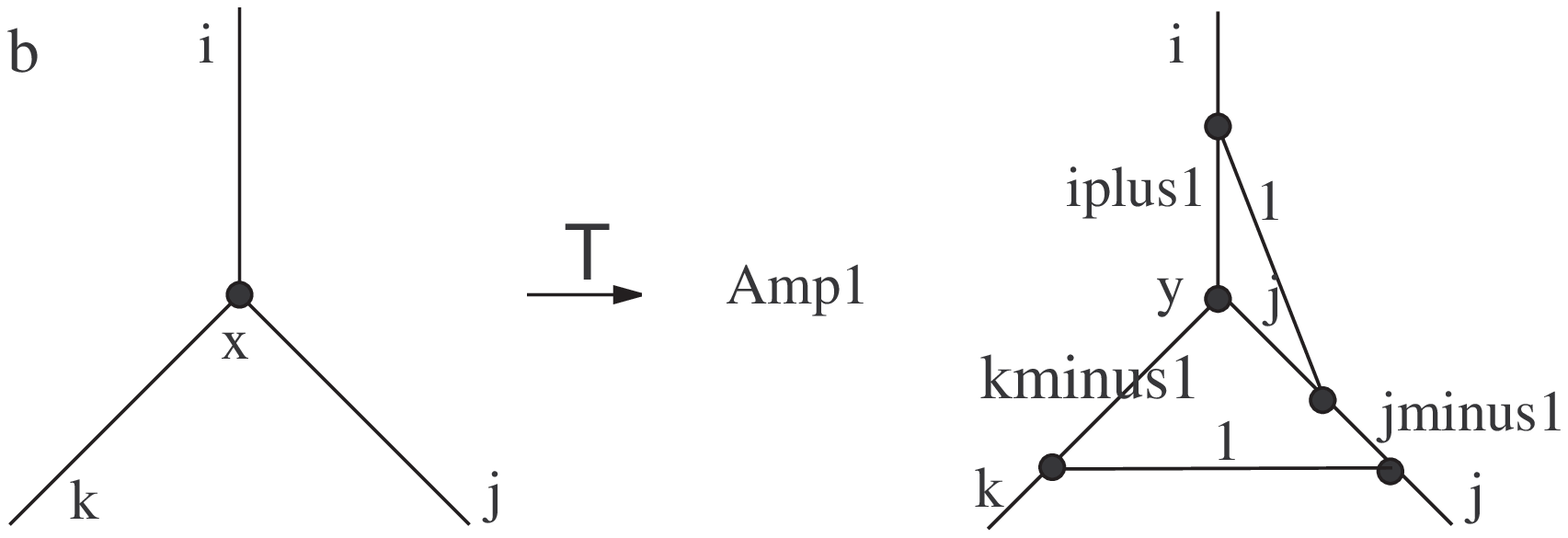}
        }
    \end{center}
    \caption{Graphical action of the HCO --- (a) shows the action Euclidean HCO
    and (b) that of $\hat{T}[N]$. $A$ and $B$ are functions of the colours $i$,
    $j$ and $k$ at the vertex $x$.}
    \label{hamfig}
\end{figure}

Many versions of the Hamiltonian constraint operator (HCO) have been proposed
\cite{RovSmo94,Borissov97,Thiemann98}. We describe here only the features of
the operator defined in \cite{Thiemann98} which we shall need in the next
section. However, the results we derive there can easily be adapted to any
other regularization of the HCO. The Lorentzian HCO can be written as a sum of
two operators
\begin{equation}\label{ham}
    \hat{C}[N]=\hat{C}_E[N] + \hat{T}[N]
\end{equation}
where $\hat{C}_E[N]$ is the HCO for the Euclidean version of the theory. The
action of the Euclidean HCO can be described graphically as follows
--- the operator acts on the nodes of the spin network state. For every pair of
edges at a node, it adds a new edge between them and changes the spins of the
edges at the node. The precise location of the new edge is unimportant. The
graphical action of the second term $\hat{T}[N]$ is less well understood, but
it can heuristically be thought of adding two non-intersecting edges at every
node. These new edges can either be added between the same pair of edges or
between two different pairs of edges at the node. Figure \ref{hamfig} shows the
graphical actions of both the terms.

The non-symmetric operator defined in \cite{Thiemann98} only creates new edges.
We take here the symmetrized version\footnote{We ignore the issues regarding
the existence of a self-adjoint extension of $\hat{T}$ \cite{Thiemann98}.}
$\hat{H}[N] = \hat{H}_E[N] + \hat{T}^S[N]$ of the operator which both adds and
removes edges. We have denoted by $\hat{H}_E[N]$ and $\hat{T}^S[N]$, the
symmetric versions of $\hat{C}_E[N]$ and $\hat{T}[N]$ respectively. The matrix
elements of the symmetric HCO are given by
\begin{eqnarray}
    \bra{s'}\hat{H}[N]\ket{s} & = & \bra{s'}\hat{C}[N]\ket{s} +
    \overline{\bra{s'}\hat{C}[N]\ket{s}}\nonumber \\
    & = & \bra{s'}\hat{C}_E[N]\ket{s}+\overline{\bra{s'}\hat{C}_E[N]\ket{s}}
    + \bra{s'}\hat{T}[N]\ket{s} + \overline{\bra{s'}\hat{T}[N]\ket{s}}.
\end{eqnarray}
The action of $\hat{C}_E[N]$ can be written using the notation of
\cite{Rovelli99} as follows:
\[
\hat{C}_E[N]\ket{s} = A^\beta_i(s) \ket{s^\beta_i;i,N}
\]
where the index $i$ runs over the nodes of the spin network $\ket{s}$. and
$\beta$ runs over pairs of edges at every node and over a sign ($\pm$) for each
of those edges. There is an implied summation over both $i$ and $\beta$. The
object on the right hand side is defined by
\[
    \braket{s;i,N}{S} = N(x_{S,i}) \braket{s}{S}\
\]
where $N(x_{S,i})$ is the value of the lapse function at the point $x$ which
the position of the $i$th node in the spin network $\ket{S}$. It is important
to note that this is the only non-diffeomorphism invariant part in the
definition of the Hamiltonian constraint. There is a regularization choice to
be made about the positions of the new vertices $y$, $y'$ and $y''$ which are
created by the action of the operator. We choose that all of these are distinct
from the position of initial point $x$. This is a arbitrary choice, but is the
most general one that can be made. In an analogous fashion, we can write the
action of the second term in eq.\ (\ref{ham}) as
\[
    \hat{T}[N]\ket{s} = B^{\beta\beta'}_i(s) \ket{s^{\beta\beta'}_i;i,N}
\]
where once again the indices $\beta$ and $\beta'$ run over pairs of edges at
every node and over a sign ($\pm$) for each of those edges. The coefficients
$A$ and $B$ are functions only of the spins of the edges and can be calculated
explicitly. An interesting point to note is that with the above regularization,
the states which result from the actions of the two pieces of the HCO are in
based on different graphs. Combined with the inner product defined in eq.\
(\ref{diffip}), this implies that there do not exist states $\ket{s}$ and
$\ket{s'}$ such that both the terms of the HCO have non-zero matrix elements
between them.

\section{Sum over Histories}

We want to study the dynamics of quantum gravity in a sum over histories
approach. The key object in this formalism is the transition amplitude from one
three-geometry to another which is given by a path integral which sums over all
histories (or four-geometries in the case of GR) which connect the two given
three-geometries. As we have already seen, in canonical GR, the spatial or
three-geometry is described by (diffeomorphism invariant) $s$-knot states. In
other words, given an initial $s$-knot states $\ket{s'}$ and a final one
$\ket{s}$ we would like to calculate:
\begin{equation}\label{formal-prop}
    A[s, s'] = \sum_\mathrm{quantum\ 4-geometries\ G} A_\mathrm{G},
\end{equation}
where the sum is over all (suitably defined) quantum four-geometries
(histories) which interpolate between the initial and final states and
$A_\mathrm{G}$ is the amplitude for each history.

Since the transition amplitude is composed of many histories, we can consider
the history to be a more elementary concept than the amplitude. Each history
can be thought of as a quantum four-geometry, hence, it is reasonable that
causality should be incorporated into each history before we calculate the
transition amplitude by performing the sum over histories. Then, each history
should have an interpretation as a quantum Lorentzian space-time with a well
defined causal structure. In the rest of the section, we will see how some of
these formal notions can be made more precise.

\subsection{The Propagator}

The finite evolution operator which propagates a state from an initial spatial
hypersurface $\Sigma_i$ a time $t=0$ to the final spatial hypersurface
$\Sigma_f$ at time $t=1$ is got by exponentiating the smeared Hamiltonian
constraint of eq.\ (\ref{ham-evol}) and integrating over the lapse function,
\begin{equation}\label{prop}
    K = \int [DN] \exp(-i\int dt \int d^3x\, N(x,t)H(x)),
\end{equation}
where $[DN]$ is the (formal) measure over the lapse. In our calculations, we
use the normalized measure on scalar functions which was constructed in
\cite{Rovelli99}. The key properties of this measure are that it is well
defined for cylindrical functions of the lapse (i.e. functions which depend
only on the value of the lapse at a finite number of points) and that it is
invariant under the action of spatial diffeomorphisms. The matrix elements of
$K$ between two three-geometries define the transition amplitudes between them.

    We can rewrite eq.\ (\ref{prop}) in the simpler form
\[
    K = \int [DT] \exp(-i\int d^3x\, T(x)H(x)),
\]
where $T(x)$ is proper time elapsed between the initial and final spatial
slices and is given by
\[
    T(x) = \int_{t=0}^1 dt\, N(x,t).
\]
In \cite{Teitel83}, the requirement that causality be incorporated into the
propagator is imposed by demanding that we include in the path integral only
those histories in which $\Sigma_f$ lies to the complete future of $\Sigma_i$.
This can be incorporated into the propagator directly by restricting the range
of integration over $T(x)$ to only positive proper times. We call the
propagator evaluated with this restriction, the \emph{causal propagator}.

Before we present the details of the operator $K$, we would like to discuss the
restriction on the proper time in some more detail. Recall that in the
canonical decomposition of classical GR, the choice $N(x)=0$ is not allowed
since it leads to a degenerate space-time metric. Thus, assuming that the lapse
is a continuous function, it must either be positive everywhere or negative
everywhere. In other words, the range of $T(x)$ is divided into two disjoint
classes, $T(x)>0$ and $T(x)<0$. The crucial assumption is that the transition
amplitude corresponds to integration over only one of these. This is motivated
by the case of the relativistic point particle where a similar restriction
leads to the Feynman propagator \cite{Teitel82}. The particular choice of
positive proper time is a matter of convention.

Since we are free to calculate the propagator in any gauge, let us now fix the
gauge using the proper time gauge conditions \cite{Teitel82}. In particular,
these require $N(x,t) = N(x)$. With this condition, we have $T(x) = N(x)$. The
causal propagator can then be written as
\begin{equation}
    K = \int_{N(x)>0} [DN] \exp(-i\int d^3x N(x)H(x)).
\end{equation}
Notice that in this form, the causal propagator is formally similar to the
projector over physical states as defined in \cite{Rovelli99}. However, in that
case, the range of integration over $N(x)$ is unrestricted. We would like to
emphasize that while the projector can be defined purely at the level of
canonical gravity, the propagator is a path-integral. We shall explore the
similarities and differences between the two operators in greater detail in a
future work. For the present, we exploit the computational techniques developed
in \cite{Rovelli99} to evaluate the causal propagator.

We wish to calculate transition amplitude which is the matrix element of $K$
between two $s$-knot states, namely $\bra{s}K\ket{s'}$. Let us start by
limiting the range of integration of $N$ such that
\[
    0 < N(x) < T.
\]
This can be thought of putting an infra-red regulator. The propagator is then
recovered by taking the limit $T \goesto \infty$. We denote by $K_T$ the
propagator where the upper limit of integration for $N(x)$ is $T$. Thus, we
have
\begin{equation}\label{trans-amp}
    \bra{s}K_T\ket{s'} = \int_{0<N(x)<T} [DN] \bra{s} \exp(-i\int d^3x
    N(x)H(x))\ket{s'}.
\end{equation}
Following  \cite{Rovelli99}, we note that the above expression is formally
three-diffeomorphism invariant and we can introduce an extra integration of the
diffeomorphism group:
\begin{equation}
    \bra{s}K_T\ket{s'} = {\cal N}\int_\mathrm{Diff} [D\phi] \int_{0<N(x)<T}
    [DN] \bra{U(\phi)S} e^{-i\hat{H}[N]}\ket{s'},
\end{equation}
where $U(\phi)$ is the finite diffeomorphism corresponding to $\phi$, ${\cal
N}$ is the normalization factor introduced in the integration over Diff and
$\ket{S}$ belongs to the diffeomorphism equivalence class of $\ket{s}$. The
integration over the diffeomorphism group is needed later to take care of the
non-diffeomorphism invariant parts of the action of the HCO. We now expand the
exponential as a formal power series
\begin{equation}
    \bra{s}K_T\ket{s'} = {\cal N}\int_\mathrm{Diff} [D\phi] \int_{0<N(x)<T}
    [DN]\bra{U(\phi)S} \sum_{n=0}^\infty \frac{(-i \hat{H}[N])^n}{n!} \ket{s'}.
\end{equation}
The $n$th power of the Lorentzian HCO can be expanded as
\begin{equation}\label{n-expand}
    (\hat{H}_E[N] + \hat{T}^S[N])^n = (\hat{H}_E[N])^n +
    :\hat{H}_E[N])^{n-1}\hat{T}^S[N]: + \cdots + (\hat{T}^S[N])^n
\end{equation}
where :(\ldots): indicates all possible orderings of the operators. We can
evaluate the matrix elements for each of the $2^n$ terms the right hand side of
eq.\ (\ref{n-expand}).  Each action of the one of the $n$ operators in the term
leads to either the addition or deletion of one or two edge(s) and a
multiplication by a numerical coefficient and a factor of $N(x)$ corresponding
to the point where the operator acted. Thus, for the first term, we get
\begin{eqnarray*}
    &&{\cal N}\frac{(-i)^n}{n!}\int_\mathrm{Diff} [D\phi] \int_{0<N(x)<T} [DN]
    \bra{U(\phi)S} (\hat{H}_E[N])^n \ket{s'}\\
    & = & {\cal N}\frac{(-i)^n}{n!}\int_\mathrm{Diff} [D\phi] A_{i_1}^{\beta_1}(s')
    \cdots A_{i_1i_2\ldots i_n}^{\beta_1\beta_2\ldots \beta_n}({s'}_{i_1i_2
    \ldots i_{n-1}}^{\beta_1\beta_2\ldots \beta_{n-1}})\braket{U(\phi)S}%
    {{s'}_{i_1i_2\ldots i_n}^{\beta_1\beta_2\ldots \beta_n}} \\
    & &\mbox{} \times \int_{N(x)>0} [DN] N(x_{i_1,U(\phi)S})
    N(x_{i_2,U(\phi)S}) \cdots N(x_{i_n,U(\phi)S}) .
\end{eqnarray*}
Similar expressions can be written down for all the other terms as well.
However, they all share the following properties: (1) for a non-zero result,
the state $\ket{s'}$ must lie in the same diffeomorphism class as the state
that results after $n$ actions of the HCO on the state $\ket{s}$; (2) the
functional dependence of the expression on $N(x)$ is now cylindrical. It is
just the product of the values of the lapse at a finite number of points. The
functional integral over $N$ can be performed in a simple fashion using the
normalized measure on scalar functions constructed in \cite{Rovelli99}. It can
simply be replaced by a finite number of ordinary integrals over the values of
$N$ at the points where the operator acted. Since the measure is diffeomorphism
invariant, the integral over Diff can be performed trivially as well. Thus,
each term in the formal power series can then be calculated and is finite. In
fact, each term in the power series has the general form
\begin{equation}
    A(s',{s'}_1, \ldots, {s'}_{n-1}) \braket{s}{{s'}_n} I_n
\end{equation}
where ${s'}_k$ is a shorthand for the $s$-knot state generated after $k$
actions of the HCO (that is, either $\hat{H}_E[N]$ or $\hat{T}^S[N]$),
$A(s',{s'}_1, \ldots, {s'}_{n-1})$ is the product of the numerical coefficients
for each action and
\[
    I_n = \frac{1}{T^n} \int^T dN(x_{i_1}) N(x_{i_1}) \int^T dN(x_{i_2})
    N(x_{i_2})\cdots \int^T dN(x_{i_n})  N(x_{i_n})
\]
is the remaining integral over the lapse. To evaluate $I_n$, we need to specify
the limits of integration for the integrals. Teitelboim's original proposal is
to perform the integration for all positive lapses. However, a na{\"\i}ve
integration following \cite{Rovelli99} ignores the causal structure induced by
multiple actions of the constraint operator. We show how this should be taken
into account in evaluating the propagator.

Intuitively, the idea is as follows: Any time there are two or more actions of
the HCO, there is a possibility that a particular action takes place at a
vertex which was created or affected by a previous action. Thus, viewed from a
space-time point of view, it lies to the causal future of the other action.
This fact must be reflected in the propagator amplitude. To do this more
precisely, let us now look at the second order ($n=2$) term of the propagator
in more detail. The matrix element for this is
\begin{eqnarray}\label{quad}
    &&\frac{(-i)^2}{2!} {\cal N}\int [D\phi] \int [DN] \bra{U(\phi)S}\hat{H}[N]
    \hat{H}[N]\ket{s'} \nonumber\\
    &=&\frac{(-i)^2}{2!} {\cal N}\int [D\phi] \int [DN]
    \left(\bra{U(\phi)S}\hat{H}_E[N]\hat{H}_E[N]\ket{s'}
    + \bra{U(\phi)S}\hat{H}_E[N]\hat{T}[N]\ket{s'}\right.\nonumber\\
    &&\mbox{} + \left. \bra{U(\phi)S}\hat{T}[N]\hat{H}_E[N]\ket{s'} +
    \bra{U(\phi)S}\hat{T}[N]\hat{T}[N]\ket{s'}\right).
\end{eqnarray}
For the discussion, let us consider the first of the four terms on the right
hand side of eq.\ (\ref{quad}), but the same considerations apply to all the
four terms. Let the first action of the Euclidean HCO take place at a node $x$
of the spin network $\ket{s'}$. Let us suppose that the action is the addition
of an edge. Let $y$ be the position of the node after the action and let $y'$
and $y''$ be the positions of the nodes where the edge was added. There are two
distinct possibilities for the second action of the Hamiltonian. It can either
act at one of the three nodes $y$, $y'$ and $y''$\footnote{The HCO defined in
\cite{Thiemann98} cannot act at the vertices $y'$ and $y''$, but this can be
traced back to the choice of the volume operator in that reference. Choosing a
different volume operator fixes this issue.} or it can act at some other node
of the spin network. Let us study these two possibilities in more detail.

    If the second action takes place at a node which is one of $y$, $y'$ or
$y''$, then in a sense, it is to the future of the first one. Otherwise, it is
causally unrelated to the first action and they commute. Thus, there is a
partial order between the two actions of the Hamiltonian. We would like to
identify this partial order with a discrete notion of causality. This is
motivated by the fact that in standard quantum field theory, operators at
spatially separated points commute, whereas ones at causally related points do
not. Since, we do not have a background metric to define the causal structure,
we suggest that \emph{the commutativity of the actions of the HCO should define
causality}. In other words, non-commuting actions correspond to causally
related points whereas commuting actions correspond to space-like separated
points.

A na{\"\i}ve functional integration over positive lapses does not capture this
feature. If we are to follow Teitelboim's basic idea of incorporating causality
in each history, then we must further modify the propagator to include this
discrete causality. This is our key observation. Let us now look at one
particular way of achieving this\footnote{We would like to emphasize that it is
by no means obvious that this choice is unique.}.

The scheme we propose is the following: Since the points $y$, $y'$ and $y''$
lie to the future of $x$ (we denote this by $x \prec y$), the domain of
integration for the proper time should reflect this. This can be understood as
follows: Suppose the second action takes place at $y$. The domain of
integration of $T(y)$ can be split into two parts $0<T(y)<T(x)$ and
$T(y)>T(x)$. The first integral vanishes because the node at $y$ does not exist
for $T(y)<T(x)$. Thus, the only contribution is from the second integral. That
is, the domain of integration can be restricted to such that proper time at the
second action is greater than that at the first action if the second action
takes place at one of the nodes created by the first action. In the proper-time
gauge, this translates to a condition on the domain of integration of the
lapse. Intuitively, we can think of the points $y$, $y'$ and $y''$ as having
been created by the action of the constraint operator at $x$ and thus being in
future of this action in every history in this class. On the other hand, if the
second action takes place at some other node $z$ which is not a product of the
first action then there is no limitation on the domain of integration over the
lapses --- we integrate over strictly positive lapses at both $x$ and $z$. This
can be thought of as the analogue of time-ordering in the absence of a
background space-time metric.

We can now calculate the functional integrals over $N(x)$ for the two cases.
When the second action is causally unrelated to the first one, then we have
\begin{eqnarray*}
    &&\frac{(-i)^2}{2!} {\cal N}\int [D\phi]\int [DN] \bra{U(\phi)S}
    \hat{H}_E[N]\hat{H}_E[N]\ket{s'} \\
    \mbox{} & = & \frac{A^{\beta_1}_{i_1}(s')%
    A^{\beta_2}_{i_2}({s'}^{\beta_1}_{i_1})}{T^2}\int_0^T dN(x) N(x_{i_1})
    \int_0^T dN(x_{i_2}) N(x_{i_2})\braket{s}{{s'}_{i_1i_2}^{\beta_1\beta_2}}\\
    \mbox{} & = & \left(\frac{1}{T} \int_0^T dN\,N\right)^2 A^{\beta_1}_{i_1}(s')
    A^{\beta_2}_{i_2}({s'}^{\beta_1}_{i_1})
    \braket{s}{{s'}_{i_1i_2}^{\beta_1\beta_2}}\\
    \mbox{} & = & \frac{T^4}{4} A^{\beta_1}_{i_1}(s') A^{\beta_2}_{i_2}
    ({s'}^{\beta_1}_{i_1})\braket{s}{{s'}_{i_1i_2}^{\beta_1\beta_2}}.
\end{eqnarray*}
If the second action is to the future of the first one, we can repeat the
above calculation and we get
\begin{eqnarray*}
    &&\frac{(-i)^2}{2!} {\cal N}\int [D\phi] \int [DN] \bra{U(\phi)S}
    \hat{H}_E[N]\hat{H}_E[N]\ket{s'}\\
    \mbox{} & = & \frac{A^{\beta_1}_{i_1}(s')%
    A^{\beta_2}_{i_2}({s'}^{\beta_1})}{T^2} \int_0^T dN(x) N(x_{i_1})
    \int_{N(x_{i_1})}^T dN(x_{i_2}) N(x_{i_2})
    \braket{s}{{s'}_{i_1i_2}^{\beta_1\beta_2}}\\
    \mbox{} & = & \left( \frac{1}{T^2}\int_0^T dN_1\,N_1 \int_{N_1}^T dN_2\,N_2
    \right)A^{\beta_1}_{i_1}(s') A^{\beta_2}_{i_2}({s'}^{\beta_1}_{i_1})
    \braket{s}{{s'}_{i_1i_2}^{\beta_1\beta_2}}\\
    \mbox{} & = & \frac{T^4}{8} A^{\beta_1}_{i_1}(s')A^{\beta_2}_{i_2}
    ({s'}^{\beta_1}_{i_1})\braket{s}{{s'}_{i_1i_2}^{\beta_1\beta_2}}.
\end{eqnarray*}
$N_1$ and $N_2$ are just shorthand notations for $N(x_{i_1})$ and $N(x_{i_2})$
respectively. Notice the numerical coefficient of the two terms are different.
Note that we have made the choice that none of the nodes $y$, $y'$ and $y''$
which result from the action of the Hamiltonian at $x$ actually coincide with
$x$. As stated earlier, this is a choice in the regularization of the operator.
Since each term in eq.\ (\ref{quad}) has the form of a numerical coefficient
which is a function of the spin network state acted upon times the value of the
lapse function at the vertex where the action took place. The functional
integral over the lapse then contributes a factor of $T^2/4$ or $T^2/8$
depending upon the causal relation between the two actions.

This calculation can be extended to all the terms in the power series. For any
given set of causal relations between the vertices, the functional integral
over the lapse can be calculated as follows: Consider a set of vertices ($x_1$,
$x_2$, \ldots, $x_n$) with causal relations ``$\prec$'' between them. Then,
\begin{equation}
    I_n = \frac{1}{T^n} \int_0^T dN_1 N_1 \int^T dN_2 N_2 \cdots \int^T dN_n
    N_n = c_n(\{x_i\},\prec)\left(\frac{T}{2}\right)^n,
\end{equation}
where the lower limits of the integrals are chosen such that if $x_i \prec
x_j$, then $N_j > N_i$ is the domain of integration. The causal factor
$c_n(\{x_i\},\prec)$ is a fraction smaller than $1$ which can be expressed as a
product of local causal factors at each vertex
\begin{equation}
    c_n(\{x_i\},\prec) = \prod_{x_i} \frac{1}{1 + k_i}
\end{equation}
where $k_i$ is the number of operator actions to the future of $x_i$. In table
1, we present the values of $c_n(\{x_i\},\prec)$ for the first few orders in
the power series. We have used a diagrammatic notation to indicate the causal
relations: Two unrelated vertices are denoted by disjoint points, whereas a
causal relation is indicated by drawing an arrow from the past point to the
future point.

\begin{table}
\begin{center}
\begin{tabular}{|l|c|c||l|c|c|}
    \cline{1-6}
    $n$ & Causal Structure & $c_n(\{x_i\},\prec)$ & $n$ & Causal Structure &
    $c_n(\{x_i\},\prec)$\\
    \cline{1-6}
    2 &\includegraphics[scale=0.3,keepaspectratio,clip]{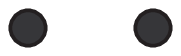}
    & 1 & 4 & \includegraphics[scale=0.3,keepaspectratio,clip]{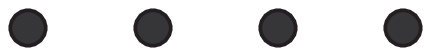}
    & 1\\
    \cline{2-3}\cline{5-6}
    & \includegraphics[scale=0.3,keepaspectratio,clip]{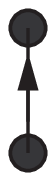} &
    \raisebox{6pt}{$\displaystyle\frac{1}{2}$} & &
    \includegraphics[scale=0.3,keepaspectratio,clip]{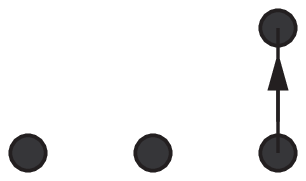} &
    \raisebox{6pt}{$\displaystyle\frac{1}{2}$}\\
    \cline{1-3}\cline{5-6}
    3 & \includegraphics[scale=0.3,keepaspectratio,clip]{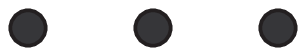}
    & 1 & & \includegraphics[scale=0.3,keepaspectratio,clip]{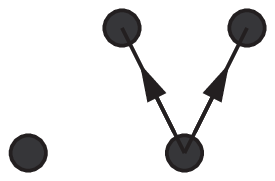} &
    \raisebox{6pt}{$\displaystyle\frac{1}{3}$}\\
    \cline{2-3}\cline{5-6}
    & \includegraphics[scale=0.3,keepaspectratio,clip]{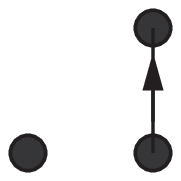} &
    \raisebox{6pt}{$\displaystyle\frac{1}{2}$} &
    & \includegraphics[scale=0.3,keepaspectratio,clip]{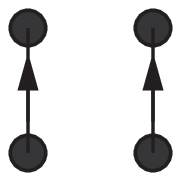} &
    \raisebox{6pt}{$\displaystyle\frac{1}{4}$}\\
    \cline{2-3}\cline{5-6}
    & \includegraphics[scale=0.3,keepaspectratio,clip]{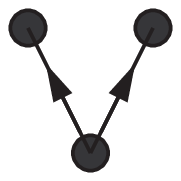} &
    \raisebox{9pt}{$\displaystyle\frac{1}{3}$} &
    & \includegraphics[scale=0.3,keepaspectratio,clip]{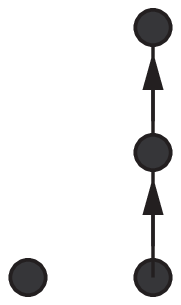} &
    \raisebox{9pt}{$\displaystyle\frac{1}{6}$}\\
    \cline{2-3}\cline{5-6}
    & \includegraphics[scale=0.3,keepaspectratio,clip]{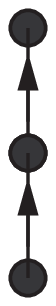} &
    \raisebox{9pt}{$\displaystyle\frac{1}{6}$} &
    & \includegraphics[scale=0.3,keepaspectratio,clip]{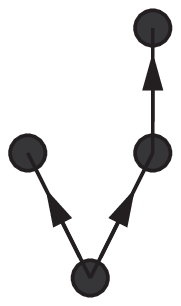} &
    \raisebox{9pt}{$\displaystyle\frac{1}{8}$}\\
    \cline{1-3}\cline{5-6}
    &&&& \includegraphics[scale=0.3,keepaspectratio,clip]{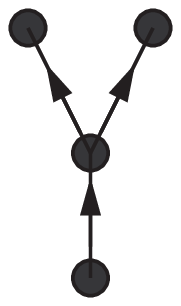} &
    \raisebox{9pt}{$\displaystyle\frac{1}{12}$}\\
    \cline{5-6}
    &&&& \includegraphics[scale=0.3,keepaspectratio,clip]{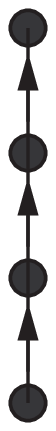} &
    \raisebox{9pt}{$\displaystyle\frac{1}{24}$}\\
    \cline{1-6}
\end{tabular}
\caption{Causal factors $c_n(\{x_i\},\prec)$ for the lowest order terms in the
power series.}
\end{center}
\end{table}

We would like to point out one very interesting feature of the causal structure
generated by multiple actions of the HCO. Namely, each vertex where the action
takes place can have at most one vertex to immediate past. This means that
causal diagrams such as
\includegraphics[height=12pt,keepaspectratio,clip]{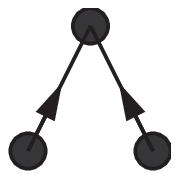} cannot appear
among the terms of the power series\footnote{This is true for the symmetric
HCO, not just the non-symmetric version.}. This is a consequence of the fact
that the action of the HCO is always localized at one vertex and can be
regarded as a manifestation of the ultra-local character of the HCO
\cite{Smolin96}. We shall return to this issue briefly in the next section.

This completes our calculation of the causal propagator except for the limit
$T\goesto\infty$. For any finite value of $T$, each term in the power series is
finite. Since the causal propagator is essentially a time-ordered exponential,
it is clear that the limit $T\goesto\infty$ should be divergent. We should
point out that this is an infra-red divergence which may be regulated by adding
a small imaginary piece to the HCO analogous to the case of the Feynman
propagator. We shall not attempt to take this limit here, but we will discuss
it some more at the end of the paper.

\subsection{Sum over Surfaces Representation}

A very powerful way of keeping track of the terms in the power series expansion
in terms of branched two-surfaces was introduced in \cite{ReisRov97}. We use
the same basic ideas, but also incorporate the causal structure which we
discussed in the previous subsection.

Each term in the power series for $\bra{s}K_T\ket{s'}$ is a sequence of
$s$-knots ${\cal F}_n = \{s, s_1, \ldots , s_n\}$ with local causal relations
between the nodes of successive spin networks in the sequence corresponding to
the action of the Hamiltonian constraint. We can represent this sequence by a
branched, colored two-surface ${\cal F}_n$ as follows:
\begin{enumerate}
\item Construct the finite two surface generated by taking the topological
product of the graph for the initial spin network with an interval.
\item Represent the first action of the Hamiltonian constraint by a branching
of the world-line of the node where the action took place.
\item Patch this onto a two surface generated by a finite piece of the
``world-sheet'' of the graph of the next spin network.
\item Repeat the above steps for all the remaining $s$-knots in the sequence.
\end{enumerate}
There is a natural local causal structure built into this surface because of
the partial order in the various actions of the Hamiltonian constraint. The
two-surface can be sliced by ``equal-time slices'' which are simply the spin
networks of the sequence. The faces in the two-surface carry SU(2) spin labels
and the lines carry intertwiners. The vertices or branching points of the two
surface correspond to the action of the HCO. Thus, the dynamics can be thought
of as being localized at the vertices. These can be thought of as being
analogous to vertices in Feynman diagrams. We refer to each of these
two-surfaces as a \emph{causal spin foam}.

\begin{figure}
    \begin{center}
        \psfrag{i1}{$x$}
        \psfrag{i2}{$y$}
        \psfrag{prec}{$x \prec y$}
        \includegraphics[height= 2.5in, keepaspectratio, clip]{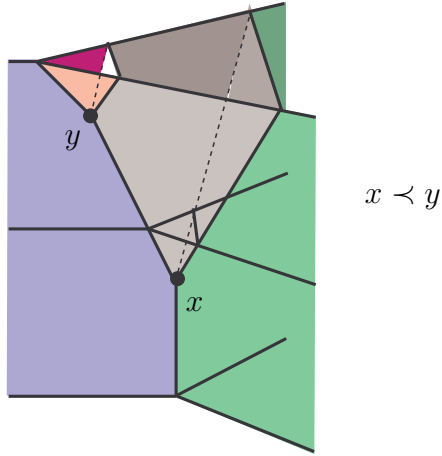}
    \end{center}
    \caption{A section of the spin foam corresponding to one of the second
    order terms in the propagator.}
\end{figure}

Each causal spin foam can be assigned the amplitude
\[
    A({\cal F}_n) = \frac{1}{n!} \prod_{v=1}^n \frac{A(v)}{1+k_v}
\]
where $A(v)$ is the numerical coefficient for the action of the HCO which takes
$s_{i-1}$ to $s_i$ and $k_v$ is the number of operator actions in the causal
future of the vertex $v$. The transition amplitude (\ref{trans-amp}) can be
then written as a sum over causal spin foams as follows:
\begin{equation}
     \bra{s}K_T\ket{s'} = \sum_{n=0}^\infty T^n \sum_{{\cal F}_n, \partial{\cal
     F}_n = s\cup s'} A({\cal F}_n).
\end{equation}
Thus, the final form of the sum over histories can be expressed as a sum over
branched, colored two-surfaces. Besides the topological information, these
two-surfaces have one additional property, namely, the causal relations between
their vertices. Based on the formal expression (\ref{formal-prop}), we can
identify these two-surfaces as being quantum, Lorentzian four-geometries. In
the next section, we give a general definition for causal spin foams, and
discuss some of their properties.

\section{Causal Spin Foams}

We can generalize the results of the previous section in a natural way. Instead
of considering two-surfaces which are generated by the action of the
Hamiltonian constraint, we consider more general branched, colored two-surfaces
and prescribe amplitudes for all possible branchings in them. This can be
thought of as an inverse prescription of the quantum Hamiltonian constraint.
For example, we can modify the ultra-local nature of the HCO by allowing more
general causal structures in the branched surfaces. The requirement that the
theory have the right classical limit, viz.\ classical general relativity, will
then choose the correct set of amplitudes. We shall now try to define the basic
structure which enables us to start addressing some of these problems. Many of
the ideas in this section are closely based on \cite{Baez98}.

From now on, we shall consider only piecewise-linear (PL) two-surfaces. The
advantage of doing so is that we can formulate the entire theory in a purely
combinatorial fashion without reference to a background manifold. We begin by
giving an intuitive discussion of what properties that these surfaces should
satisfy. Firstly, they should relate to spin networks in a natural way. This
means that they interpolate between an initial spin network to a final one.
Also, the notion of causality which arose in the previous section should be
incorporated in these two-surfaces. This causality should also give us a
prescription for ``slicing'' a two-surface and getting a spin network. We now
make these notions more precise.

Consider a oriented two-dimensional cellular complex $\kappa$. It consists of a
collection of point (0-cells), edges (1-cells) and faces (2-cells). Let
$V(\kappa)$ denote the set of vertices of the complex, $E(\kappa)$ the set of
edges and $F(\kappa)$ the set of faces. Since, we have an oriented complex, for
every edge $e \in E(\kappa)$, there exist maps $s$ and $t$ that map $e$ to its
source and target in $V(\kappa)$.

Next, we need to include causality. This is done by imposing be a partial order
$\prec$ among the elements of $V(\kappa)$. That is, for all $x,y,z \in
V(\kappa)$:
\begin{eqnarray*}
&& x \prec y\ \mathrm{and}\ y \prec z \implies x \prec z\\
&& x \prec y\ \mathrm{and}\ y \prec x \implies  x = y
\end{eqnarray*}
These properties state that the order on the set is transitive and reflexive.
Such sets are referred to as causal sets in the physics literature
\cite{Sorkin90}. Further, since each term we consider has a finite number of
vertices, the vertex set satisfies the local finiteness requirements that are
usually imposed on causal sets. In particular, the partial order $\prec$ gives
us relations between any two vertices which are joined by an edge. We shall
call an edge ``causal'' if we have the condition
\begin{equation}
s(e) \prec t(e) \hspace{0.5in}\mathrm{or}\hspace{0.5in} t(e) \prec s(e),
\end{equation}
otherwise we shall call it ``acausal'' or``space-like''. We adopt the
convention that the orientation of a causal edge is always future-directed,
that is, $s(e) \prec t(e)$.

Only faces which have at least one causal edge are non-empty (i.e., belong to
the foam). This corresponds to the intuitive notion that the ``world-sheet'' of
the edges of the spin networks should be causal surfaces. Also, every
space-like edge can belong to at most two faces of $\kappa$. This is because
the dynamics defined by the action of the HCO is localized at vertices. We
would like to retain this feature. Further, we assign to every face $f$ in
$\kappa$ which has at least one causal edge, a representation $j_f$ of SU(2)
and to every edge which belongs to at least two faces we shall assign an
intertwiner $I$ which defines the map
\[
    I: \bigotimes_{f_i\in \mathrm{incoming\ faces}} j_{f_i}
    \longrightarrow \bigotimes_{f_o\in \mathrm{outcoming\ faces}} j_{f_o}.
\]
A face is considered to be incoming if the orientation it induces on the edge
agrees with the intrinsic orientation of the edge and considered outgoing
otherwise. This in particular means that the faces which meet at space-like
edges have the same spin and the intertwiner associated with such an edge is
identity. The causal spin foam ${\cal F}_C$ is then defined as the object
$(\kappa,\prec,\vec{\jmath},\vec{I})$.

    We started by requiring that a causal spin foam give us an history
between some initial and some final spin network. This translates into a
requirement on the boundary of ${\cal F}_C$. The boundary of $\kappa$ consists
of all edges $e_{\del\kappa}$ which lie in only one face and the vertices that
form the endpoints of these. Since the foam is supposed to be the interpolation
between two spatial spin networks, we need to specify further properties of the
boundary $\del\kappa$ of $\kappa$ such that this is satisfied. It is sufficient
to require that the set of vertices $V_{\del\kappa}$ which belong to the
boundary of $\kappa$ consists of two disjoint sets $V_{\gamma_1}$ and
$V_{\gamma_2}$ such that for any $v \in V_{\gamma_1}$ there is some $u \in
V_{\gamma_2}$ such that $v \prec u$ and that there are no causal relations
between any two elements of each of the sets $V_{\gamma_1}$ and $V_{\gamma_2}$.
$V_{\gamma_1}$ is then the set of vertices of the graph $\gamma_1$ on which the
initial spin network is based. The edges $e_{\del\kappa}$ are labeled by the
same representation of the group as the face they border, while the vertices in
$V_\del\kappa$ are labeled by the intertwiners which labeled the edges which
end on them. Hence, a causal spin foam ${\cal F}_C$ can be regarded as a
mapping from an initial spin network to a final one.

Intermediate spin networks should correspond to spatial slices of the foam. Any
acausal, connected, closed subset of the 1-skeleton of $\kappa$ (i.e. the set
of all 0-cells and 1-cells which belong to $\kappa$) can be chosen as a graph
on which a spin network is based. Since all space-like edges have at most two
faces incident on them, we assign the representation $j$ associated with either
of those faces to the edge. Generically, we can also assign the intertwiner of
the causal edge passing through every vertex to the vertex itself. Thus, in
general, this will give us a spin network. We cannot slice through vertices
where more than two causal edges meet as there is no unambiguous definition of
the intertwiner for these vertices. Slicing to the past and future of these
vertices gives us different spin networks. Thus these are the ``Feynman
vertices'' of the theory where the dynamics is effectively localized.

This completes our characterization of a causal spin foam. To summarize the key
points, we give a definition:

\begin{defn}
A causal spin foam $\cal F_C = \{\kappa,\prec,\vec{\jmath},\vec{I}\}$ is a map
from a spin network $s_1 = \{\gamma_1,\vec{\jmath}_1,\vec{I}_1\}$ to a spin
network $s_2 = \{\gamma_2,\vec{\jmath}_2,\vec{I}_2\}$ such that
\begin{enumerate}\itemsep=-\parskip
    \item $\kappa$ is an oriented two-dimensional cellular complex;
    \item $\prec$ is a partial order on the vertex set $V(\kappa)$;
    \item an edge of $\kappa$ is called causal if its endpoints are related by
    $\prec$, otherwise it called acausal;
    \item every face $f\in F(\kappa)$ has at least one causal edge;
    \item every face of $\kappa$ is labeled by a representation $j_f$ of SU(2);
    \item every edge of $\kappa$ is labeled by an intertwining tensor $I_e$
    such that
    \[
        I_e: \bigotimes_{f_i\in \mathrm{incoming\ faces}} j_{f_i}
    \longrightarrow \bigotimes_{f_o\in \mathrm{outcoming\ faces}} j_{f_o};
    \]
    \item the boundary of $\kappa$, $\partial\kappa = \gamma_1\cup\gamma_2$
    and for every vertex $v\in V(\kappa\cap\gamma_1), \exists u\in V(\kappa
    \cap\gamma_2)$ such that $v\prec u$ and $\gamma_i$ are acausal sets;
    \item the spins $\vec{\jmath}_i$ and intertwiners $\vec{I}_i$ of $s_i$ agree
    with those on the faces and edges of $\cal F_C$ which intersect the
    boundary.
\end{enumerate}
\end{defn}

Dynamics can be considered as the assignment of amplitudes $A({\cal F}_C)$ to
each foam. In particular, this amplitude can be written as
\begin{equation}
    A({\cal F}_C) = \prod_{v \in {\cal F}_C} A_c(v)
\end{equation}
where the amplitude assigned to each vertex is a local quantity, that is, it
depends only upon the spins and intertwiners in the neighborhood of the vertex.
Based on the result for the previous section, this amplitude should also
contain a factor which depends upon the causal structure of the vertex. A
specification of the amplitude $A_c(v)$ for all possible vertices in the theory
can be regarded as being equivalent to giving the definition of the HCO.

In the next section, we give an example of a specific causal spin foam model
which does not arise from any Hamiltonian constraint.

\section{The causal evolution model as a causal spin foam.}

The causal evolution model \cite{Marko97} can be viewed as causal spin foam
model. We shall start by giving a brief review of our understanding of the
essential features of this model.

Start with a (abstract) three-dimensional simplicial complex ${}^3\!\Delta$ in
which every face is labeled by a spin $j$ (representation of SU(2)) and every
tetrahedron is labeled by an intertwiner which defines the map from
$\displaystyle{\otimes_\mathrm{faces}} j$ to the trivial representation. This
is supposed to correspond to a spatial slice. The dual one-skeleton of the
complex ${}^3\!\Delta$ is a four-valent spin network (the model restricts to
spin networks based on four-valent graphs). Now the amplitude between an
initial state (${}^3\!\Delta_1$ and a final state (${}^3\!\Delta_2$) can be
written as
\begin{equation}
{\cal A}_{{}^3\!\Delta_1 \longrightarrow {}^3\!\Delta_2} = \sum_\mathrm{all\
4-simplicial\ complexes\ {}^4\!\Delta} \left(\prod_\mathrm{4-simplices} {\cal
A}_\mathrm{4-simplex}\right),
\end{equation}
where we consider all the four-dimensional simplicial complexes ${}^4\!\Delta$
which have as their boundaries ${}^3\!\Delta_1 \cup {}^3\!\Delta_2$. Each
four-simplex in ${}^4\!\Delta$ consists of 5 tetrahedra and 10 faces. The
causal evolution model further specifies that the 4-simplex has a particular
causal structure, that is, $k$ of the tetrahedra lie in the past of the
remaining $(5-k)$. Thus, the amplitude $\cal A$ which corresponds to every
4-simplex is a function in the space $\mathrm{Inv}(j_1\otimes
j_2\otimes\cdots\otimes j_{10})$ which reflects this causal structure.

The contact with spin networks was made in the case of the three-simplex by
looking at the dual one-skeleton of ${}^3\!\Delta$. We may expect that the
causal spin foam may be recovered by considering the dual two-skeleton of the
four-complex.

Let us consider one four-simplex in the interior of a particular
${}^4\!\Delta$. The dual two-skeleton can be constructed by assigning a 0-cell
to the 4-simplex itself, a 1-cell to every tetrahedron in the 4-simplex and a
2-cell to every face of the 4-simplex. By assigning the labels of every element
in the simplex to its dual element, we can generate a 2-dimensional cellular
complex whose faces are labeled by spins and whose edges are labeled by
intertwiners. The only things that remains to be checked is that every face in
the dual two-skeleton has a time like edge in its boundary. To verify this, we
observe that the edges in the dual 1-skeleton correspond to joining the centers
of neighboring 4-simplices. Since, these 4-simplices have causal relations
between them, the dual edges are obviously causal. (In the language of causal
evolution, there is an order in which we placed the 4-simplices to generate
${}^4\!\Delta$ and this is identified as a causal order).

There is certain amount of subtlety associated with taking the duals of the
three-complexes that form the boundary. The way to define this is to first
construct the dual one-skeleton of the boundary 3-complex by associating an
edge to every face in it. Next, construct the duals for these faces in the
four-dimensional sense, considering only the parts of these surfaces which are
inside the four-complex. These will then be bounded by the one-skeleton of the
boundary. Thus, the boundary of the dual two-skeleton is in fact the union of
the graphs on which the initial and final spin networks are based.

Thus, the causal evolution model satisfies all the conditions in Definition 1.
In fact, the causal evolution model can be viewed as a causal spin foam model
in which only two-surfaces which are dual 2-skeletons of a 4-complex and the
causal structures which are local to each 4-simplex are included in the sum
over histories. It is worth noting that the causal evolution model also gives a
method for constructing (a limited class of) causal spin foams.

\section{Discussion}

In this paper, we have shown two main results. We have constructed the causal
propagator in canonical quantum gravity as a formal power series. This was made
possible by the use of spin network states and the existence of a
diffeomorphism invariant measure for the lapse function. We found that it was
necessary to modify the original proposal in \cite{Teitel83} to include the
discrete causality that was introduced by multiple actions of the HCO. In fact,
we used the non-commutativity of the HCO to define a causal structure. This was
our first key result.

The other key result is that the power series for the propagator can be
represented as a sum over colored, branched two-surfaces which we call causal
spin foams. We gave a general definition for a causal spin foam and showed that
the causal evolution model \cite{Marko97} can be regarded as a causal spin foam
model.

We shall conclude this paper with a discussion of some of the open issues that
remain to be understood. These are listed below in no particular order.
\begin{itemize}
\item The propagator which we have considered is motivated by studying the
Feynman propagator for the point particle. However, the role that it plays in
quantum  gravity is not completely understood. In particular, its relation to
the projector onto physical states needs to be clarified.

\item We have presented a particular concrete proposal to incorporate the
causal structure which results from the multiple actions of the HCO. The
four-diffeomorphism invariance of this proposal deserves to be better
understood. In particular, we have suggested, that the non-commutativity of the
HCO defines the causal structure of the quantum space-time. This should be
studied in greater detail by coupling the gravity to matter fields and
verifying that our definition of causality is consistent with that in usual
quantum field theory.

Also, we have concentrated on the range on integration for the lapse. The
dependence on the measure chosen for the lapse is unclear. We should emphasize
though, that the discrete causal structure will determine the range of
integration for the lapse, once the measure is chosen.

\item The limit $T\goesto\infty$ needs further study. There are two different
avenues which can be explored in this regard. As we suggested, it may be
possible to regulate this in a fashion analogous to the case of the Feynman
propagator for the relativistic point particle. Another proposal is that this
limit should only be taken in physically relevant quantities where it may make
sense \cite{Rovelli99}.

\item The details of the calculation in section 3 depend on a particular
regularization of the HCO. However, given any other regularizations of the
operator, we can repeat the calculation for them. However, we would like to
point out that we gave an example of a model which can be viewed as an inverse
definition of the constraint operator. The causal spin foam models and the
Hamiltonian constraint should be viewed as two sides of the same problem --- a
better understanding of one will help in the study of the other.

\item The appearance of causal spin foams in the path integral for Lorentzian
quantum gravity indicates that they may be thought of as quantum Lorentzian
space-times. This aspect needs to studied in greater detail. In particular, the
relation of this work to \cite{Barrett99}, where the authors try to construct a
spin foam model for Lorentzian gravity based on the representation theory of
the Lorentz group, requires further investigation. It is interesting to note
that in three dimensions, a path integral model based on the group SO(2,1) can
be constructed in which each term has a consistent causal structure
\cite{Freidel99}.

\item A related issue, which arises in a natural fashion, is the choice of
causal spin foams that should be included in the sum over histories. It may be
that it is sufficient to limit the sum to a subset of all causal spin foams
(for example, the ones which arise in the causal evolution model). It may even
turn out that this question is irrelevant in the classical limit --- i.e.,
there may be more than one class of models for which classical GR is
recovered\footnote{For an approach to defining the HCO based on requiring the
correct classical limit, see \cite{Gupta99}}. Further work is needed before
this question can be answered in a definite way.
\end{itemize}

This work adds to the understanding of the role that causality plays in the
dynamics of quantum gravity. However, many issues are still unclear and further
work needs to be done before answers to physical questions can be extracted.

\section*{Acknowledgements}

I would like to thank Lee Smolin, Fotini Markopoulou, Laurent Freidel and
Michael Reisenberger for numerous helpful discussions and suggestions. This
work was supported in part by NSF grant PHY-9514240 to the Pennsylvania State
University, a Miller fellowship from the Physics Department, and a gift from
the Jesse Phillips Foundation.

\end{document}